\documentclass[%
 reprint,
 amsmath,amssymb,
  pra,
]{revtex4-1}

\usepackage{graphicx}
\usepackage{dcolumn}
\usepackage{bm}

\begin{document}


\title{Interaction of Diatomic Molecules with Photon Angular Momentum}

\author{Thomas B. Bahder}
\affiliation{%
Aviation and Missile Research, 
      Development, and Engineering Center, \\    
US Army RDECOM, 
Redstone Arsenal, AL 35898, 
U.S.A.}%

\date{\today}

\begin{abstract}
The interaction of a diatomic molecule with  photons carrying well-defined angular momentum and parity is investigated to determine whether photon absorption can induce molecular  rotational transitions between states having angular momentum $\Delta J >1$.  A transformation from laboratory coordinates to coordinates with origin at the center-of-mass of the nuclei is used to obtain the interaction between the photons and the molecule's center-of-mass, electronic, and rotational degrees of freedom.  For molecules making transitions between rotational levels, there is a small parameter, $ k a \ll 1 $, where $k$ is the photon wave vector and $a$ is the size of the molecule, which enters into the $Ej$ and $Mj$ photon absorption probabilities.  For electric photons having arbitrary angular momentum $j \hbar$, the probability  of absorbing an $E(j+1)$ photon divided by the probability of absorbing an $Ej$ photon, scales as $ (k a)^2 /( 2 j+1)^2$. The probability of absorbing an $Mj$ photon, divided by the probability of absorbing and $Ej$ photon scales according to the same factor.
\end{abstract}

\maketitle


\section{\label{Intro}Introduction}
A molecule is significantly more complicated than an atom because, in addition to center of mass motion, a molecule has vibrational and rotational degrees of freedom.
Work on molecular spectra has a long history.  A significant milestone was made in 1927 by Born and Oppenheimer~\cite{Born1927}, who  assumed that the lighter electrons in a molecule adjust adiabatically to the slower motion of the heavier nuclei, which remain at all times in their instantaneous ground state.   Further progress in understanding molecular spectra was made by treating a rotating molecule as a symmetric top or asymmetric top~\cite{Wang1929}, sometimes called a rigid rotor~\cite{LL_QuantumMechanics,Davydov_QuantumMechanics}, see Ref.~\cite{VanVleck1929} and references contained therein.  In the rigid rotor treatment of molecules, vibrational and center-of-mass motion are essentially neglected.

Solving the complete quantum mechanical many body problem consisting of electrons and nuclei for a given type of molecule will in-principle provide the energy spectrum of the molecule.  In practice, this is a complicated task which usually results in heavy numerical work.  For this reason, much analytic work has been done on simple diatomic molecules~\cite{LL_QuantumMechanics,Davydov_QuantumMechanics}.  For modern reviews on the theory of the molecular Hamiltonians, see Meyer~\cite{Meyer2002}, and for an overview about diatomic molecules, see Brown and  Carrington~\cite{Brown2003}.

While the spectrum of a molecule can in-principle be obtained by solving the corresponding quantum mechanical many body problem, in experiments we usually probe molecules with electromagnetic fields.  In order to interpret experiments, we must calculate selection rules and probabilities of transition between molecular quantum states.  This involves investigating the coupling of a molecule to the electromagnetic field.  The fundamental theory of a second quantized electromagnetic radiation field coupled to charges is discussed in Ref.~\cite{AB_QuantumElectrodynamics,LL_QuantumElectrodynamics}.  Detailed treatments dealing with radiation coupling to molecules are given by Atkins and Wooley~\cite{Atkins1970}, Woolley~\cite{Woolley1971,Woolley1975},  Craig and Thirunamachandran~\cite{Craig_MolecularQuantumElectrodynamics} and a recent detailed treatment is given by  Sindelka~\cite{Sindelka2006}. 

In the past two decades, a completely different line of investigation has dealt with the angular momentum carried by the electromagnetic field.   Beth~\cite{Beth1936} made the first experimental observation that light carries angular momentum. Beth demonstrated that if a beam of right-hand circularly polarized light passed through a birefringent plate, and the beam was converted into a left-hand polarized beam, the plate experienced a torque of $2 \hbar$ for each photon in the beam.    This experiment is cited as a demonstration that a light beam carries spin angular momentum.  Years later, in 1992, Allen et al. showed that laser light with a Laguerre-Gaussian amplitude distribution has well-defined orbital angular momentum~\cite{Allen1992}.  Many theoretical and experimental investigations followed, see Ref.~\cite{TwistedPhotonsBook}.  For paraxial beams, i.e., in the paraxial approximation, it has been demonstrated that the total angular momentum can be separated into a sum of spin angular momentum and orbital angular momentum~\cite{Allen1999}.  However, for a general electromagnetic radiation field, spin and orbital angular momentum cannot be separately defined, and only the total angular momentum is a well-defined quantity~\cite{AB_QuantumElectrodynamics,Davydov_QuantumMechanics,LL_QuantumElectrodynamics}.   The interaction of atoms with light carrying angular momentum were explored theoretically in a number of works~\cite{Huang1994,Jauregui2004,Grinter2008}.

Within the rigid rotor model of a molecule, one can easily see that an electromagnetic radiation field  can induce (dipole, quadrupole, etc...)  rotational transitions in a diatomic molecule through its multipole moments,  see for example p. 601 in Section 136 of Davydov\cite{Davydov_QuantumMechanics}. However, the rigid rotor treatment neglects the molecule center-of-mass degree of freedom.  In the absence of a radiation field, the molecular center-of-mass motion decouples from the other degrees of freedom and there is no issue to consider further, see Eqs.~(\ref{KE_inCMofMolecule}) and (\ref{kineticenergyNucCoords}).  However, in the presence of a radiation field, the molecular center of mass plays a key role in the molecule-field interaction, see Eq.~(\ref{InteractionHamCMofNucleiiCoords}) and Ref.~\cite{Thomas1970,Thomas1970a,Thomas1970b,Thomas1971,Sindelka2006}.  References~\cite{Thomas1970,Thomas1970a,Thomas1970b,Thomas1971} discuss the non-trivial role of the center of molecular mass when a classical plane polarized electromagnetic wave  (corresponding to photons having $j = \hbar$)  interacts with a molecule.  
In order to investigate in detail molecular absorption of photons having arbitrary angular momentum ($j>n \hbar)$, where $n>1$, one has to revisit the interaction Hamiltonian between the molecule and radiation field.  As already remarked, this is complicated because a molecule has several degrees of freedom: center-of-mass motion, electronic motion,  vibrational motion, and rotational motion.  

In this manuscript, I consider a simple diatomic molecule interacting with photons carrying arbitrary angular momentum, $j \hbar$. I investigate the mechanism that leads to  transitions between molecular rotational levels, with emphasis on the role of the molecular center-of-mass. Of particular interest are probabilities of transitions between rotational levels whose angular momentum differs by  more than one unit, $\Delta J  >1$.  I write down a  Hamiltonian for a diatomic molecule, for the radiation field, and for their interaction.  By making successive transformations from the laboratory coordinates into the center-of-mass coordinates, and then to center-of-molecule coordinates, I obtain interaction terms that show the coupling of a diatomic molecule to the radiation field. For the molecule, there is a small parameter, $k a \ll 1$, where $k$ is the wave vector and $a$ is the size of the molecule. I find how the probability of the molecule absorbing an $Ej$ and $Mj$ multipole photon scales with $k a$.   In this treatment, I neglect all relativistic interactions containing electron and nuclear spins.      

\section{\label{Hamiltonian}Molecular and Field Hamiltonian}
I consider a closed system consisting of a diatomic molecule interacting with a quantized radiation field.  The Hamiltonian can be described as a sum of three terms~\cite{Davydov_QuantumMechanics,LL_QuantumElectrodynamics,AB_QuantumElectrodynamics,Atkins1970,Woolley1971,Woolley1975}:
\begin{equation}
{{\hat H}} = {\hat H}_{molecule} + {{\hat H}_{field}} + {\hat H}_{\mathop{\rm int}} (t)
\label{TotalHamiltonian}
\end{equation}
where ${\hat H}_{molecule}$ is the Hamiltonian for the molecule alone (with no field present), ${{\hat H}_{field}}$ is the Hamiltonian for the field (with no molecule present) and ${\hat H}_{\mathop{\rm int}}$ is the interaction  Hamiltonian between the radiation field and the molecule.   I assume the diatomic molecule consists of $n$ electrons and two nuclei.  I use a laboratory system of Cartesian coordinates, ${\bf R}_i= (X_i, Y_i,Z_i)$, $i = 1,2, \ldots,n$, for the electrons and ${\bf R}_\alpha=(X_\alpha, Y_\alpha,Z_\alpha)$, $\alpha=1,2$, for the nuclei. I take the basis vectors of the Cartesian coordinate system to be $({\bf e}_x,{\bf e}_y,{\bf e}_z )\equiv({\bf e}_1,{\bf e}_2,{\bf e}_3 )$. The position of the origin of the laboratory system of coordinates is arbitrary. In what follows, I distinguish between coordinates, e.g., coordinates of the $i$th electron, ${\bf R}_i= (X_i, Y_i,Z_i)$, and the geometric object,  $(X_i  {\bf e}_x,  Y_i  {\bf e}_y,  Z_i  {\bf e}_z,)$, which represents the position of the $i$th electron and is independent of the coordinate system.  Note that these two quantities, ${\bf R}_i= (X_i, Y_i, Z_i)$ and   $(X_i  {\bf e}_x,  Y_i  {\bf e}_y,  Z_i  {\bf e}_z,)$, transform differently under change of coordinate systems.  The first one transforms as coordinate components and the second is a geometric quantity that does not transform.

The Hamiltonian for the diatomic molecule can be written as~\cite{AB_QuantumElectrodynamics,Meyer2002}
\begin{equation}
{{\hat H}_{molecule}} = \hat T + {{\hat V}_{e - e}} + {{\hat V}_{n - n}} + {{\hat V}_{e - n}}
\label{MoleculeHamiltonian}
\end{equation}
The kinetic energy operator for electrons and nuclei is given by
\begin{equation}
\hat T = \sum\limits_{i = 1}^n {\frac{{{\bf{\hat P}}_i^2}}{{2m}}}  + \sum\limits_{\alpha  = 1,2} {\frac{{{\bf{\hat P}}_\alpha ^2}}{{2{M_\alpha }}}} 
\label{KineticEnergyLabFrame}
\end{equation}
where $m$ is the mass of an electron and $M_1$ and $M_2$ are the masses of the nuclei, and $n$ is the number of electrons.   Note that I have not assumed that the molecule is electrically neutral. The electron and nuclear momentum operators, ${\bf{\hat P}}_i  = \frac{\partial }{{\partial {{\bf{R}}_i}}}$ and ${\bf{\hat P}}_\alpha  = \frac{\partial }{{\partial {{\bf{R}}_\alpha }}}$, are conjugate to the coordinates, 
${\bf R}_i= (X_i, Y_i,Z_i)$ and ${\bf R}_\alpha=(X_\alpha, Y_\alpha,Z_\alpha)$. The Coulomb interaction terms for electrons, nuclei, and electrons and nuclei, respectively, are given by
\begin{equation}
\begin{array}{l}
 {{\hat V}_{e - e}} = \frac{{\,{e^2}}}{{4\pi {\varepsilon _0}}}\sum\limits_{i < j}^{} {\frac{1}{{\left| {{{\bf{R}}_i} - {{\bf{R}}_j}} \right|}}}  \\ 
 {{\hat V}_{n - n}} = \frac{{{Z_1}\,{Z_2}\,{e^2}}}{{4\pi {\varepsilon _0}}}\frac{1}{{\left| {{{\bf{R}}_2} - {{\bf{R}}_1}} \right|}} \\ 
 {{\hat V}_{e - n}} =  - \frac{{\,{e^2}}}{{4\pi {\varepsilon _0}}}\sum\limits_{\alpha  = 1,2} {\sum\limits_{i = 1}^n {\frac{{{Z_\alpha }}}{{\left| {{{\bf{R}}_i} - {{\bf{R}}_\alpha }} \right|}}} }  \\ 
 \end{array}
\label{CoulombInteractions}
\end{equation}
where $e$ is the electronic charge,  $\varepsilon _0$ is the permitivity of vacuum, $Z_\alpha$ is the atomic number and $M_\alpha$ is the mass of nucleus $\alpha$.

\section{\label{PhotonField}Photons with Well-Defined Angular Momentum}

For the the radiation field, I can choose Laguerre-Gaussian modes~\cite{Allen1999,Romero2002}, which are solutions of the paraxial wave equation, and in which a separation can be made between spin angular momentum and orbital angular momentum. These types of photons carry angular momentum $\hbar l$.  Such modes are convenient when dealing with cylindrical symmetry, however, they are only approximate solutions of Maxwell equations.  Alternatively, we can choose vector spherical harmonic modes~\cite{Davydov_QuantumMechanics,LL_QuantumElectrodynamics}, which are {\it exact} solutions of Maxwell's equations, but then spin angular momentum and orbital angular momentum cannot be separated. This is not a flaw of vector spherical harmonics, rather it is a general feature of electromagnetic radiation fields. The vector spherical harmonic modes are a complete set of vector eigenfunctions, so any field can be expressed in terms of these modes. For these reasons, I take the radiation field to be  described by vector spherical harmonic modes.   The Hamiltonian for the radiation field, see Eq.~(\ref{TotalHamiltonian}),  is given by
\begin{equation}
{{\hat H}_{field}} = \sum\limits_{\lambda ,k,j,m} {\hbar {\omega _k}\left( {{{\hat a}^\dag }_{\lambda kjm}\,{{\hat a}_{\lambda kjm}} + \frac{1}{2}} \right)\,} 
\label{FieldHamiltonian}
\end{equation}
where ${{{\hat a}^\dag }_{\lambda kjm}}$ and ${{{\hat a}}_{\lambda kjm}}$ are the creation and destruction operators for electric or magnetic multipole photons, labeled by $\lambda=(E,M)$, respectively, having wave vector magnitude $k= \omega_k/c$, and carrying angular momentum $j \hbar$ and angular momentum projection $m \hbar$ on z-axis .   The creation and annihilation operators satisfy the commutation relations~\cite{Davydov_QuantumMechanics} 
\begin{equation}
\left[ {{{\hat a}_{\lambda kjm}},{{\hat a}^\dag }_{\lambda '{\kern 1pt} k'{\kern 1pt} j'{\kern 1pt} m'}} \right] = {\delta _{\lambda {\kern 1pt} \lambda '}}{\delta _{k{\kern 1pt} k'}}\,{\delta _{j{\kern 1pt} j'}}\,{\delta _{m{\kern 1pt} m'}}
\label{CommuttationRelations}
\end{equation}
Each field state with well-defined photon number has well-defined energy.  Each photon in a state labeled by quantum numbers ($\lambda, k, j, m$) has wave vector $k$, square of angular momentum $j(j+1) \hbar^2$, z-component of angular momentum $m \hbar$, and parity $(-1)^{j}$ for $\lambda = E$ and parity  $(-1)^{j+1}$ for $\lambda = M$. Photons are of two types: an $Ej$ photon is an electric photon and a $Mj$ photon is a magnetic photon, each carrying $j \hbar$ units of angular momentum~\cite{Davydov_QuantumMechanics,AB_QuantumElectrodynamics,LL_QuantumElectrodynamics}.  The vector potential operator is given by
\begin{equation}
{\bf{\hat A}}({\bf{r}},t) = \sum\limits_{\lambda ,k,j,m} {\left[ {{{\hat a}_{\lambda kjm}}\,{\bf{A}}_{kjm}^\lambda ({\bf{r}},t) + {{\hat a}^\dag }_{\lambda kjm}\,{\bf{A}}_{kjm}^{\lambda \; * }({\bf{r}},t)} \right]}
\label{VectorPotentialOperator}
\end{equation}
where ${{\bf{A}}_{kjm}^E ({\bf{r}},t)}$ is the photon wave function for electric 2$^j$ pole radiation and  ${{\bf{A}}_{kjm}^M ({\bf{r}},t)}$ is the photon wave function for magnetic 2$^j$ pole radiation~\cite{Davydov_QuantumMechanics,AB_QuantumElectrodynamics,LL_QuantumElectrodynamics}.   In a large sphere of radius $\cal{ R}$, the magnetic and electric photon wave functions are given by~\cite{Davydov_QuantumMechanics}
\begin{equation}
{\bf{A}}_{kjm}^M({\bf{r}},t) = \frac{1}{{\sqrt V }}{f_j}(kr)\,{\bf{Y}}_{jjm}(\theta ,\phi )\,{e^{ - i{\omega _k}t}}
\label{PhotonMagneticWaveFunction}
\end{equation}
and
\begin{eqnarray}
{\bf{A}}_{kjm}^E({\bf{r}},t) &  =  & \frac{1}{\sqrt{V} \,\sqrt{2j + 1} }  \left[ \sqrt{j} \, f_{j + 1}(kr) {\bf{Y}}_{j,j + 1,m}(\theta ,\phi ) \right. \nonumber \\
                  &  &  \left. -  \sqrt{j + 1} \,{f_{j - 1}}(kr){\bf{Y}}_{j,j - 1,m}(\theta ,\phi ) \right]\,{e^{ - i{\omega _k}t}}
\label{PhotonElectricWaveFunction}
\end{eqnarray}
where photon frequency is $\omega_k=c k$ and the spherical quantization volume is $V$.  The radial functions, $f_j(kr)$, are given in terms of Bessel functions of the first kind~\cite{Davydov_QuantumMechanics}, $J_{\nu + \frac{1}{2}}(x)$
\begin{equation}
{f_l}(k \,r) = \frac{{{{\left( {2V} \right)}^{1/2}}}}{R}\frac{1}{{{J_{l + \frac{3}{2}}}\left( {{x_{l + \frac{1}{2},n}}} \right)}}\frac{{{J_{l + 
\frac{1}{2}}}\left( {k \,r} \right)}}{{{r^{1/2}}}}
\label{RadialFunctions}
\end{equation}
and the allowed set of discrete wave vectors, $k_{ln}$, where $l$ and $n$ are positive integers, are given by
\begin{equation}
{J_{l + \frac{1}{2}}}\left( {{k_{l{\kern 1pt} n}}\, \cal{R}} \right) = 0
\label{DiscreteKs}
\end{equation}
These discrete wave vectors determine the photon energies, $\hbar \omega_k = \hbar c k_{l \, n}$, in a sphere of radius $R$.  For a given value of $l$, the spectrum of allowed wave vectors, $\{ k_{l \, n} \}$ is the same for electric and magnetic photons.

The three vector wave functions (vector spherical harmonics), ${\bf Y}_{j,l,m}(\theta ,\phi )$,  \mbox{$j=l+1, l, |l-1|$}, are defined as direct products of two different irreducible representations of the three-dimensional rotation group, ${\bf e}_\mu$,  and $Y_{lm}(\theta ,\phi ) $, coupled by Clebsch-Gordon coeficients:
\begin{equation}
{{\bf{Y}}_{j,l,m}}(\theta ,\phi )   =    \sum\limits_\mu  {\left\langle {{1,l,\mu ,m - \mu }}  
 \mathrel{\left | {\vphantom {{1,l,\mu ,m - \mu } {jm}}}  \right. \kern-\nulldelimiterspace}
 {{jm}} \right\rangle } \,{{\bf{e}}_\mu }\,{Y_{l,m - \mu }}
\label{YvecDef}
\end{equation}
Here ${\bf e}_\mu $ are three spin $S=1$ eigenfunctions, $\chi_\mu (\sigma)$,  whose components are  ${\bf e}_\mu = \{ \chi_\mu (+1), \chi_\mu (0), \chi_\mu (-1) \}$,  and $\sigma= {+1,0,-1}$ is the spin variable~\cite{Davydov_QuantumMechanics,LL_QuantumMechanics}.  The  ${\bf e}_\mu$, $\mu=\{-1,0,+1\}$, are the spherical basis vectors, which are simultaneous eigenfunctions of the spin angular momentum operators $\hat{{\bf S}}^2$ and $\hat{S}_z$:
\begin{eqnarray}
 \hat{{\bf S}}^2 \,{\bf e}_\mu  & =   & 2{\kern 1pt} {\bf e}_\mu  \nonumber \\  
 \hat{S}_z \,{\bf e}_\mu        &  =  &\mu \,{\bf e}_\mu  
\label{SpinEigenfunction}
\end{eqnarray}
The spherical basis vectors are related to the Cartesian basis vectors, $\{ {\bf e}_x, {\bf e}_y, {\bf e}_z \}$, by
\begin{eqnarray}
 {{\bf{e}}_ + } &  =  &  - \frac{1}{{\sqrt 2 }}\left( {{{\bf{e}}_x} + i{{\bf{e}}_y}} \right) \nonumber \\ 
 {{\bf{e}}_ - } &  = & \frac{1}{{\sqrt 2 }}\left( {{{\bf{e}}_x} - i{{\bf{e}}_y}} \right) \nonumber \\ 
 {{\bf{e}}_0}    & =  &  {{\bf{e}}_z}  
\label{Cartesian-SphericalBasis}
\end{eqnarray}
In terms of the Cartesian basis vectors,  $\{ {\bf e}_x, {\bf e}_y, {\bf e}_z \}$, the action of the spin operator is given by
\begin{equation}
\hat{S}_i {\bf e}_k  = i \sum_{l=1,2,3} \, e_{ikl} \, {\bf e}_l
\label{SpinOpActingOnBasisVector}
\end{equation}
where $e_{ikl}$ is the totally antisymmetric Levi-Civita tensor, where $e_{123}=+1$.

The $Y_{l,m }$ in Eq.~(\ref{YvecDef}) are the usual scalar spherical Harmonics that are eigenfunctions of the orbital angular momentum operators $\hat{{\bf L}}^2$ and $\hat{L}_z$:
\begin{eqnarray}
 \hat{{\bf L}}^2 \, Y_{lm}  & =   & l(l+1)  Y_{lm}   \nonumber \\  
 \hat{L}_z \,Y_{lm}        &  =  & m Y_{lm}  
\label{OrbitalAMeigenfunction}
\end{eqnarray}
According to  the definitions in Eq.(\ref{YvecDef})--(\ref{OrbitalAMeigenfunction}), the product functions,
${{\bf{e}}_\mu }\,{Y_{l,m }}$, are simultaneous eigenfunctions of  $\hat{{\bf L}}^2$, $\hat{L}_z$,  $\hat{{\bf S}}^2$ and $\hat{S}_z$, with eigenvalues $l(l+1)$, $m$, $2$, and $\mu$, respectively.

The spherical vector harmonics, ${{\bf{Y}}_{j,l,m}}(\theta ,\phi )$, form an irreducible representaion of the three-dimensional rotation group on the unit sphere, and are simultaneous eigenfunctions of  $\hat{J}^2$,  $\hat{J}_z$,   $\hat{{\bf L}}^2$, and  $\hat{{\bf S}}^2$:
\begin{eqnarray}
{{\hat J}_z}{{\bf{Y}}_{j,l,m}}\left( {\theta ,\phi } \right) & = & m{{\bf{Y}}_{j,l,m}}\left( {\theta ,\phi } \right) \\
\label{Jzeq}
{{\hat J}^2}{{\bf{Y}}_{j,l,m}}\left( {\theta ,\phi } \right) & = & j(j + 1){{\bf{Y}}_{j,l,m}}\left( {\theta ,\phi } \right)\\
\label{j2eqn}
{{\hat L}^2}{{\bf{Y}}_{j,l,m}}\left( {\theta ,\phi } \right) & = & l(l + 1){{\bf{Y}}_{j,l,m}}\left( {\theta ,\phi } \right)\\
\label{L2eqn}
{{\hat S}^2}{{\bf{Y}}_{j,l,m}}\left( {\theta ,\phi } \right) & = & 2{{\bf{Y}}_{j,l,m}}\left( {\theta ,\phi } \right)
\label{S2eqm}
\end{eqnarray}

The  vector spherical harmonics, ${{\bf{Y}}_{j,l,m}}(\theta ,\phi )$, satisfy the orthogonality relation
\begin{equation}
\int {{d^2}\Omega } \,{\bf{Y}}_{jlm}^ * \left( {\theta ,\phi } \right) \cdot {\bf{Y}}_{j'l'm'}^{}\left( {\theta ,\phi } \right) = {\delta _{j\,j'}}\,{\delta _{l\,l'}}\,{\delta _{m\,m'}}
\label{Yorthogonality}
\end{equation}
where the integration is over the unit sphere, and completeness relation
\begin{equation}
\sum\limits_{j\,l\,m} {{{\bf{Y}}_{j,l,m}}\left( {\theta ,\phi } \right){{\bf{Y}}_{j,l,m}}\left( {\theta ',\phi '} \right)}  = {\bf{\mathord{\buildrel{\lower3pt\hbox{$\scriptscriptstyle\leftrightarrow$}} 
\over I} }}\,{\delta ^2}\left( {\Omega  - \Omega '} \right)
\label{completeness}
\end{equation}

Under the parity operator (spatial inversion), the basis vectors ${\bf e}_\mu$ change sign, and we take the scalar spherical harmonics, $Y_{lm}$, to be defined so that  they are multiplied by $(-1)^l$, and therefore the vector spherical harmonics ${\bf Y}_{j,l,m}(\theta ,\phi )$ are eigenfunctions of parity with eigenvalues $(-1)^{l+1}$.

From Eq.~(\ref{PhotonMagneticWaveFunction}) we see that magnetic photons have well-defined spin and orbital angular momentum, since ${\bf{A}}_{kjm}^M({\bf{r}},t) $ is an eigenfunction of spin and orbital angular momentum.  However, from  Eq.~(\ref{PhotonElectricWaveFunction}) we see that electric photons with quantum numbers $(j,m)$ are a superposition of orbital angular momentum $l=j-1$ and $l=j+1$, and do not have well-defined orbital angular momentum.

The number operator for photons of type $\lambda$ is given by
\begin{equation}
\hat{{n}}_\lambda = \sum\limits_{k,j,m} {{{\hat a}^\dag }_{\lambda kjm}\,{{\hat a}_{\lambda kjm}}\,} 
\label{NumberOperator}
\end{equation}

\section{\label{InteractionHamiltonian}Molecule-Field Interaction}
The electromagnetic field couples to the charges in the molecule, which consist of electrons and nuclei. The interaction Hamiltonian describing  the coupling of the radiation field to the molecule is given by~\cite{Atkins1970,Woolley1971,Woolley1975,LL_QuantumElectrodynamics,Meyer2002,Sindelka2006} 
\begin{equation}
{{\hat H}_{{\mathop{\rm int}} }} (t) = \frac{e}{m}\sum\limits_{i = 1}^n {{\bf{\hat A}}({{\bf{R}}_i},t) \cdot {{{\bf{\hat P}}}_i}}  - e\sum\limits_{\alpha  = 1}^2 {\frac{{{Z_\alpha }}}{{{M_\alpha }}}{\bf{\hat A}}({{\bf{R}}_\alpha },t) \cdot {{{\bf{\hat P}}}_\alpha }} 
\label{InteractionHamiltonian}
\end{equation}  
where the vector potential operator is in laboratory coordinates, and is given in Eq.~(\ref{VectorPotentialOperator}).
In Eq.~(\ref{InteractionHamiltonian}), the first term is the coupling of the field to electrons and the second term is the coupling of the field to the nuclei.  The radiation field, ${\bf{\hat A}}({\bf r},t )$, is quantized in the laboratory coordinates, ${\bf R}$. The electron and nuclear momentum operators in Eq.~(\ref{InteractionHamiltonian}), ${\bf P}_i$ and ${\bf P}_\alpha$, are also in laboratory coordinates. I have dropped terms that are quadratic in the vector potential because they are responsible for photon scattering and are small for photon absorption and emission.  Also, I have dropped small relativistic terms that couple the electron and nuclear spins to the field~\cite{LL_QuantumMechanics,Davydov_QuantumMechanics,LL_QuantumElectrodynamics}.  Furthermore, in Eq.~(\ref{InteractionHamiltonian}), I have assumed the Coulomb gauge for the vector potential, so that ${\rm div}\,\, {\bf{\hat A}}({\bf r},t ) = 0$ and the scalar potential is zero.  

In many experiments, the wavelength of the electromagnetic radiation field is long compared to the dimensions of the physical system, such as  the molecule.  We can make a Taylor series expansion of the vector potential terms  
${\bf {\hat A}}({\bf R}_i,t)$ and ${\bf{\hat A}}({\bf R}_\alpha, t) $ about the center of mass of the molecule ${\bf R}_o$.  Keeping only the first term, I  make the approximations
\begin{equation}
{\bf{\hat A}}({{\bf{R}}_i},t) \approx {\bf{\hat A}}({{\bf{R}}_\alpha },t) \approx {\bf{\hat A}}({{\bf{R}}_o},t)
\label{LongWaveApproximation}
\end{equation}

\section{\label{TransformationToMoleculeCM}Transformation to Center of Mass of the Molecule}
 
In order to explore the transfer of angular momentum from photons to molecules, we must look at degrees of freedom that can ''absorb" angular momentum.  To this end, I transform the molecular Hamiltonian in Eq.~(\ref{MoleculeHamiltonian}) from the laboratory coordinates to center-of-mass-of-molecule coordinates.  I take the Cartesian basis vectors in laboratory coordinates to be collinear with the basis vectors in center-of-mass-of-molecule coordinates.  I follow the notation used by Brown and Carrington~\cite{Brown2003}.  The position of the center of mass of the molecule, in laboratory coordinates, is given by
\begin{equation}
{{\bf{R}}_o} = \frac{m}{M}\sum\limits_{i = 1}^n {{{\bf{R}}_i} + \frac{1}{M}\sum\limits_{\alpha  = 1}^2 {{M_\alpha }{{\bf{R}}_\alpha }} } 
\label{CMofMolecule}
\end{equation}
where the total molecular mass is given by
\begin{equation}
M = \sum\limits_{i = 1}^n {m_i} + \sum\limits_{\alpha  = 1}^2 {{M_\alpha }}  = n m + M_1 + M_2
\label{TotalMAss}
\end{equation}
where $n$ is the number of electrons in the molecule.  
The transformation of electron and nuclear coordinates, from laboratory coordinates, ${{\bf{R}}_i}$ and $ {{\bf{R}}_\alpha } $,   to the center of mass of molecule coordinates, ${{{\bf{R'}}}_i}$ and ${{{\bf{R'}}}_\alpha }$, is given by
\begin{eqnarray}
 {{{\bf{R'}}}_i} & = & {{\bf{R}}_i} - {{\bf{R}}_o}     \nonumber \\ 
  {{{\bf{R'}}}_\alpha } & = & {{\bf{R}}_\alpha } - {{\bf{R}}_o}   
\label{CMofMoleculeTransformation}
\end{eqnarray}
Note that the transformations in Eq.~(\ref{CMofMoleculeTransformation}) are transformations of vector components, see comment above Eq.~(\ref{MoleculeHamiltonian}). I introduce the inter-nuclear vector, ${\bf R}$, in terms of the coordinates of the two nuclei (in laboratory coordinates)
\begin{equation}
{\bf R} = {\bf R}_2 - {\bf R}_1
\label{InternuclearVector}
\end{equation}
Using the above definitions of the transformation to the center-of-mass-of-molecule coordinates,  the transformation of the kinetic energy of the molecule, given in Eq.~(\ref{KineticEnergyLabFrame}), to the center-of-mass-of-molecule coordinates, gives~\cite{Brown2003}
\begin{equation}
T = \frac{1}{{2M}}{\bf{P}}_o^2 + \frac{1}{{2\mu }}{\bf{P}}_{\bf{R}}^2 + \frac{1}{{2m}}\sum\limits_{i = 1}^n {{\bf{P'}}_i^2 - } \frac{1}{{2M}}\sum\limits_{j,k = 1}^n {{{{\bf{P'}}}_i} \cdot } {{{\bf{P'}}}_k}
\label{KE_inCMofMolecule}
\end{equation}
where the momentum operators
\begin{equation}
{{\bf{P}}_o} = \frac{\hbar }{i}\frac{\partial }{{\partial {{\bf{R}}_o}}}
\label{CenterOfMassMomentum}
\end{equation}
\begin{equation}
{{\bf{P}}_{\bf{R}}} = \frac{\hbar }{i}\frac{\partial }{{\partial {\bf{R}}}}
\label{RelativeMomentum}
\end{equation}
\begin{equation}
{{\bf{P}}_j^\prime} = \frac{\hbar }{i}\frac{\partial }{{\partial {{\bf{R}}_j^\prime}}}
\label{ElectronMomentum}
\end{equation}
are conjugate to the coordinates, 
${\bf R}_o$, ${\bf R}$   and ${\bf R}_j^\prime $, and where $\mu$ is the reduced nuclear mass 
\begin{equation}
\frac{1}{\mu } = \frac{1}{{{M_1}}} + \frac{1}{{{M_2}}}
\label{ReducedNuclearMass}
\end{equation}
The first term in Eq.~(\ref{KE_inCMofMolecule}) is the center of mass kinetic energy of the molecule, expressed in laboratory coordinates. The second term in Eq.~(\ref{KE_inCMofMolecule}) is the relative kinetic energy of the nuclei, expressed in terms of the inter-nuclear vector ${\bf R}$. The third term in Eq.~(\ref{KE_inCMofMolecule}) is the kinetic energy of the electrons.  The last term in Eq.~(\ref{KE_inCMofMolecule}) is the so-called mass polarization terms and results from fluctuations of electron position about the massive nuclei.

The transformation of the Coulomb interaction terms, given in Eq.~(\ref{CoulombInteractions}), from the laboratory coordinates to center-of-mass-of-molecule coordinates, is done similarly.

\section{\label{TransformationToNucleiCM}Transformation to Center of Mass of the Nuclei}
 
In order to discuss rotation of the molecule about the line joining the nuclei, I define the position of the center of mass of the nuclei (in center-of-mass-of-molecule  coordinates):
\begin{equation}
{{{\bf{R'}}}_N} = \frac{1}{{{M_1} + {M_2}}}\left( {{M_1}{{{\bf{R'}}}_1} + {M_2}{{{\bf{R'}}}_2}} \right)
\label{NuclearCM}
\end{equation}

The positions of the electrons in the center-of-mass-of-nuceli coordinates are given by
\begin{eqnarray}
{{{\bf{R''}}}_i} & = & {{{\bf{R'}}}_i} - {{{\bf{R'}}}_N}  \nonumber \\
   & = & {{{\bf{R'}}}_i} + \frac{1}{{{M_1} + {M_2}}}\sum\limits_{j = 1}^n {{m_j}} {{{\bf{R'}}}_j}
\label{electroncoordinatesNucFrame}
\end{eqnarray}

In the center-of-mass-of-nuceli coordinates, the momentum operator conjugate to the electron position, ${{{\bf{R''}}}_i}$, is given by
\begin{equation}
{{{\bf{P''}}}_j} = \frac{\hbar }{i}\frac{\partial }{{\partial {{{\bf{R''}}}_j}}}
\label{ElectronMomemntumCMofNucleiiCoords}
\end{equation}

The transformation of nuclear coordinates, ${\bf R}_\alpha$, to center-of-mass-of-nuceli coordinates is similar.

The total kinetic energy of the molecule, given in equation Eq.~(\ref{KE_inCMofMolecule}), transformed to center-of-nuclei coordinates, is given by~\cite{Judd1975,Brown2003}
\begin{eqnarray}
T & = & \frac{1}{2M}{\bf{P}}_o^2 + \frac{1}{{2\mu }}  {\bf{P}}_{\bf{R}}^2 + \frac{1}{{2m}}\sum\limits_{i = 1}^n {\bf{P''}}_i^2 \nonumber \\
  & + &  \frac{1} {{2\left( {{M_1} + {M_2}} \right)}}\sum\limits_{j,k = 1}^n {{{{\bf{P''}}}_j} \cdot } {{{\bf{P''}}}_k}
\label{kineticenergyNucCoords}
\end{eqnarray}

Transforming the Coulomb terms in Eq.~(\ref{MoleculeHamiltonian}) and (\ref{CoulombInteractions}) to the center-of-mass-of-the-nuclei coordinates (which are designated by two primed superscripts), these terms become 
\begin{equation}
\begin{array}{l}
{{\hat V}_{e - e}} = \frac{{\,{e^2}}}{{4\pi {\varepsilon _0}}}\sum\limits_{i < j}^{} {\frac{1}{{\left| {{{{\bf{R''}}}_i}_j} \right|}}}   \\ 
{{\hat V}_{n - n}} = \frac{{{Z_1}\,{Z_2}\,{e^2}}}{{4\pi {\varepsilon _0}}}  \frac{1}{R}       \\ 
{{\hat V}_{e - n}} =  - \frac{{\,{e^2}}}{{4\pi {\varepsilon _0}}}\sum\limits_{\alpha  = 1,2} {\sum\limits_{i = 1}^n {\frac{{{Z_\alpha }}}{{\left| {{{{\bf{R''}}}_{i\alpha }}} \right|}}} }   \\ 
 \end{array}
\label{CoulombInteractionsCMofNucleii}
\end{equation}
where $|{{{\bf R''}}}_{ij}| = \left| {{{\bf R''}}}_i - {{{\bf R''}}}_j \right|$ is the distance between electrons, 
$|{{{\bf R''}}}_{i\alpha} | = \left| {{{{\bf R''}}}_i} - {{{\bf R''}}}_\alpha  \right|$ is the distance between the $i$th electron and $\alpha$th nucleus, and ${\bf R}$ is the inter-nuclear vector defined in Eq.~(\ref{InternuclearVector}), with magnitude $R={\bf R}$. Note that the ${{{\bf R''}}}_{i \alpha}$ are expressible in terms of  the electron coordinates, ${{{\bf R''}}}_i$, and the inter-nuclear vector,  ${\bf R}$, so the molecular wavefunction depends on the center of mass coordinate, ${\bf R}_o$, the inter-nuclear coordinate, ${\bf R}$, and the electron coordinates ${{{\bf R''}}}_i$, for $i=1,2,\cdots,n$. 

In summary, when the molecule Hamiltonian in laboratory coordinates, given in Eq.~(\ref{MoleculeHamiltonian}), is transformed into the center-of-mass-of-nuceli coordinates, it is given by the sum of  Eq.~(\ref{kineticenergyNucCoords}) and (\ref{CoulombInteractionsCMofNucleii}). Note that in the absence of the electromagnetic field, the center-of-mass motion of the molecule decouples from the internal motion of the molecule, i.e., the molecular Hamiltonian does not contain mixed terms with center-of-mass and internal degrees of freedom. So, in the absence of the radiation field,  the total molecular wavefunction is the product: 
\begin{equation}
{\psi _{QLM}}\left( {{{\bf{R}}_o}} \right)\,\phi \left( {{\bf{R}},{{{\bf{R''}}}_1},{{{\bf{R''}}}_2}, \cdots ,{{{\bf{R''}}}_n},} \right)
\label{ProductWaveFunction}
\end{equation}
where the center-of-mass wavefunction is of the form
\begin{equation}
{\psi _{QLM}}\left( {{{\bf{R}}_o}} \right) = {R_{QL}}\left( {Q{R_o}} \right)\,{Y_{LM}}\left( {{\theta _o},{\phi _o}} \right)
\label{CMwavefunction}
\end{equation} 
 and $\phi \left( {{\bf{R}},{{{\bf{R''}}}_1},{{{\bf{R''}}}_2}, \cdots ,{{{\bf{R''}}}_n},} \right)$ is the wave function for internal degrees of freedom.

\section{\label{MoleculeFieldInteactionInCMofNuclei}Interaction Hamiltonian in CM of Nuclei Coordinates}

As previously stated, in the absence of the radiation field, the molecule is described by a Hamiltonian that is the sum of  Eq.~(\ref{kineticenergyNucCoords}) and (\ref{CoulombInteractionsCMofNucleii}).  In this case, the center of mass motion, described by coordinate ${\bf R}_o$,  decouples from the internal motion, which is described by coordinate ${\bf R}$.    However, when an external radiation field is present, see the interaction Hamiltonian given in Eq.~(\ref{InteractionHamCMofNucleiiCoords}),   the radiation field introduces a coupling between the center of mass motion of the molecule and its internal motions~\cite{Sindelka2006}.

Next, I assume that the molecule interacts with radiation, according to the interaction Hamiltonian in Eq.~(\ref{InteractionHamiltonian}).  It is possible to re-write the molecule-field interaction in a form that allows more insight than that given by the interaction Hamiltonian in Eq.~(\ref{InteractionHamiltonian}). In order to do this, I make the long  wavelength approximation, given in Eq.~(\ref{LongWaveApproximation}), in the interaction Hamiltonian in Eq.~(\ref{InteractionHamiltonian}).  Then, I transform the electron and nuclear momentum operators to center-of-mass-of-nuclei coordinates. The molecule-field interaction Hamiltonian ${\hat{H}_{{\mathop{\rm int}} }}(t)$ in Eq.~(\ref{InteractionHamiltonian}) then becomes
\begin{eqnarray}
{\hat{H}_{{\mathop{\rm int}} }} (t) &  = & \frac{e}{M}(n - {Z_1} - {Z_2})\,{\bf{\hat A}}({{\bf{R}}_o},t) \cdot \,{{{\bf{\hat P}}}_o} +  \nonumber \\ 
                          &  + & {\kern 1pt} e\left( {\frac{1}{m} + \frac{{{Z_1} + {Z_2}}}{{{M_1} + {M_2}}}} \right){\kern 1pt} {\bf{\hat A}}({{\bf{R}}_o},t) \cdot \,\sum\limits_{i = 1}^n {{{{\bf{P''}}}_i}}  \nonumber \\ 
                          &  + & {\kern 1pt} e\left( {\frac{{{Z_1}}}{{{M_1}}} - \frac{{{Z_2}}}{{{M_2}}}} \right){\kern 1pt} {\bf{\hat A}}({{\bf{R}}_o},t) \cdot \,{{{\bf{\hat P}}}_R} 
\label{InteractionHamCMofNucleiiCoords}
\end{eqnarray}
where ${{{\bf{\hat P}}}_o}$ is the relative momentum operator for the center of mass of the molecule (in laboratory coordinates) that is defined in Eq.~(\ref{CenterOfMassMomentum}), ${{{{\bf{P''}}}_i}} $ is the momentum operator of the $i$th electron (in center-of-mass-of-nuclei coordinates) given in Eq.~(\ref{ElectronMomemntumCMofNucleiiCoords}), and ${{{\bf{\hat P}}}_R}$ is the momentum operator that is conjugate to the inter-nuclear vector ${\bf R}$ and is given by Eq.~(\ref{InternuclearVector}).

The total Hamiltonian for the molecule, the field, and their interaction is  given by the sum of  Eq.~(\ref{kineticenergyNucCoords}), (\ref{CoulombInteractionsCMofNucleii}) and (\ref{InteractionHamCMofNucleiiCoords}).  The coordinates in the total Hamiltonian are the center-of-mass position (in laboratory coordinates), ${\bf R}_o$, the inter-nuclear position (nuclear relative coordinate) ${\bf R}$, and the electron positions in the center-of-mass-of-nuclei coordinates, ${{\bf{R''}}_i}$.  The basis states for the combined system of molecule and field are of the form
\begin{equation}
\Psi \left( {{{\bf{R}}_o},{\bf{R}},{{{\bf{R''}}}_1}, \ldots ,{{{\bf{R''}}}_n}} \right)\,\left| {{{({n_1})}_{{\alpha _1}}}\,,{{({n_2})}_{{\alpha _2}}}\,, \cdots } \right\rangle 
\label{wavefunction}
\end{equation}
where $\Psi \left( {{{\bf{R}}_o},{\bf{R}},{{{\bf{R''}}}_1}, \ldots ,{{{\bf{R''}}}_n}} \right)$ are the molecular states that are eigenstates of the sum of  Eq.~(\ref{kineticenergyNucCoords}) and (\ref{CoulombInteractionsCMofNucleii}), and  $\left| {{{({n_1})}_{{\alpha _1}}}\,,{{({n_2})}_{{\alpha _2}}}\,, \cdots } \right\rangle$ are the photon states that are eigenstates of Eq.~(\ref{FieldHamiltonian}), where there are $n_i$ photons in mode $\alpha _i$ and I use the short-hand notation $\alpha=\{\lambda, k, j, m  \}$.

The first term in the interaction Hamiltonian in Eq.~(\ref{InteractionHamCMofNucleiiCoords}) is an ionic term, and is proportional to the pre-factor  $n - {Z_1} - {Z_2}$. This pre-factor is zero for neutral molecules, whose electron number is equal to the sum of the atomic numbers of the two nuclei, $n = {Z_1} + {Z_2}$.  For the case of a molecule that is not neutral (an ion) this first term couples the momentum of the center of mass of the molecule to the radiation field.  This term leads to center of mass motion of the (charged) molecule as a whole.

The second term in the interaction Hamiltonian in Eq.~(\ref{InteractionHamCMofNucleiiCoords}) couples the sum of electron momenta (in center-of-mass-of-nuclei coordinates) to the radiation field. This term allows a transfer of angular momentum from the radiation field to the electrons.  The pre-factor for this second term is never zero.

The third term in the interaction Hamiltonian in Eq.~(\ref{InteractionHamCMofNucleiiCoords}) couples the momentum, 
\begin{equation}
{{\bf{P}}_{\bf{R}}} =  - i\hbar \left[ {\frac{\partial }{{\partial {R_X}}}{{\bf{e}}_x} + \frac{\partial }{{\partial {R_Y}}}{{\bf{e}}_y} + \frac{\partial }{{\partial {R_Z}}}{{\bf{e}}_z}} \right]
\label{P_R}
\end{equation}
which is conjugate to the inter-nuclear vector, \mbox{${\bf{R}} = \left( {{R_X},{R_Y},{R_Z}} \right)$}, to the electromagnetic radiation field, ${\bf{\hat A}}({{\bf{R}}_o},t)$. Here, $({\bf e}_x, {\bf e}_y, {\bf e}_z )$ are the basis vectors for the space-fixed coordinates of \mbox{${\bf{R}} = \left( {{R_X},{R_Y},{R_Z}} \right)$}.   The pre-factor for this term is zero for homo-nuclear diatomic molecules~\cite{Sindelka2006}, which have $M_1 =M_2$ and $Z_1=Z_2$. Homo-nuclear diatomic molecules have an inversion symmetry about the nuclear center of mass.   Molecules which have $Z_1=Z_2$ but $M_1 \ne M_2$, (e.g., different isotopes, such as HD),  lack this inversion symmetry, and then this third term in Eq.~(\ref{InteractionHamCMofNucleiiCoords}) is generally non-zero.  This third term contains a coupling of the radiation field to the rotation of the molecule about the line joining the nuclei.  
This can be seen by considering the unitary  transformation, $\cal{M}$, from space-fixed coordinates with origin at the center-of-nuclear mass, \mbox{${\bf{X}} = \left( {{X},{Y},{Z}} \right)$}, to molecule-fixed coordinates, \mbox{${\bf{x}} = \left( {{x},{y},{z}} \right)$}, where the $z$-axis is pointing along the inter-molecular vector ${\bf R}_2 -{\bf R}_1$:

\begin{equation}
\left( {\begin{array}{*{20}{c}}
   X  \\
   Y  \\
   Z  \\
\end{array}} \right) = \cal{M} \left( {\phi ,\theta ,\chi } \right) \cdot \left[ {\begin{array}{*{20}{c}}
   x  \\
   y  \\
   z  \\
\end{array}} \right]
\label{transformation}
\end{equation}
where $\left( {\phi , \theta ,\chi } \right)$ are the Euler angles between coordinates \mbox{${\bf{X}} = \left( {{X},{Y},{Z}} \right)$} and \mbox{${\bf{x}} = \left( {{x},{y},{z}} \right)$}.  I take the sequence of rotations through angles $\{\phi ,\theta ,\chi \}$ to be along the $z$, $y$, $z$ axes, so the rotation matrix is given by~\cite{Judd1975,Zare1988,Brown2003}  
\begin{widetext}
\begin{equation}
\cal{M} \left( {\phi , \theta ,\chi } \right) = \left( {\begin{array}{*{20}{c}}
   {\cos \phi \cos \theta \cos \chi  - \sin \phi \sin \chi } & { - \sin \phi \cos \chi  - \cos \phi \cos \theta \sin \chi } & {\cos \phi \sin \theta }  \\
   {\sin \phi \cos \theta \cos \chi  + \cos \phi \sin \chi } & {\cos \phi \cos \chi  - \sin \phi \cos \theta \sin \chi } & {\sin \phi \sin \theta }  \\
   { - \sin \theta \cos \chi } & {\sin \theta \sin \chi } & {\cos \theta }  \\
\end{array}} \right)
\label{RotationMatrix}
\end{equation}
\end{widetext}

For internuclear vector $ {\bf R}= \left( {{R_X},{R_Y},{R_Z}} \right)$, the transformation from space-fixed coordinates to molecule-fixed coordinates, \mbox{$\left( {{R_X},{R_Y},{R_Z}} \right) \rightarrow \left( {R,\theta ,\phi ,\chi } \right)$}, is given by
\begin{equation}
\left( {\begin{array}{*{20}{c}}
   {{R_X}}  \\
   {{R_Y}}  \\
   {{R_Z}}  \\
\end{array}} \right) = \cal{M}   \left( {\phi , \theta ,\chi } \right) \cdot \left[ {\begin{array}{*{20}{c}}
   0  \\
   0  \\
   R  \\
\end{array}} \right]
\label{transformation_SpaceFixed--Molecule-Fixed}
\end{equation}
For a diatomic molecule, since the vector \mbox{${\bf{R}} = \left( {{R_X},{R_Y},{R_Z}} \right)$} is parallel to the molecule-fixed $z$-axis, the components $ \left( {{R_X},{R_Y},{R_Z}} \right)$ do not depend on $\chi$, and we are free to choose its value for convenience~\cite{Judd1975,Zare1988,Brown2003}.  Using the transformation in Eq.~(\ref{transformation_SpaceFixed--Molecule-Fixed}),  the relative momemtum in Eq.~(\ref{P_R}) becomes~\cite{Brown2003}:
\begin{equation}
{{\bf{P}}_{\bf{R}}} = \hbar \left[ { - \frac{1}{R}{\bf{n}} \times {\bf{N}} - i{\bf{n}}\frac{\partial }{{\partial R}}} \right]
\label{RelativeMomentum}
\end{equation}
where ${\bf n}$ is the inter-nuclear unit vector, given in space-fixed coordinates by
\begin{equation}
{\bf{n}} = \frac{{\bf R}_2 -{\bf R}_1}{|{\bf R}_2 -{\bf R}_1|}= \sin \theta \,\cos \phi \,{{\bf{e}}_x} + \sin \theta \,\sin \phi \,{{\bf{e}}_y} + \cos \theta \,\,{{\bf{e}}_z}
\label{interNuclearVector}
\end{equation}
and ${\bf N}$ is the angular momentum operator in space-fixed coordinates with origin at center-of-mass of the nuclei
\begin{eqnarray}
 {\bf{N}}   &  =  & i\left[ {\cot \theta \cos \phi \frac{\partial }{{\partial \phi }} + \sin \phi \frac{\partial }{{\partial \theta }} - {\mathop{\rm c}\nolimits} \sec \theta \cos \phi \frac{\partial }{{\partial \chi }}} \right]{{\bf{e}}_x}  \nonumber \\ 
  & + &  i\left[ {\cot \theta \sin \phi \frac{\partial }{{\partial \phi }} - \cos \phi \frac{\partial }{{\partial \theta }} - {\mathop{\rm c}\nolimits} \sec \theta \sin \phi \frac{\partial }{{\partial \chi }}} \right]{{\bf{e}}_y} \nonumber  \\ 
  & - & i\frac{\partial }{{\partial \phi }}{{\bf{e}}_z} 
\label{NvectorComponents}
\end{eqnarray}

Using the form of ${\bf P}_{\bf R}$ in Eq.~(\ref{RelativeMomentum}) in the third term in Eq.~(\ref{InteractionHamCMofNucleiiCoords}), gives 
\begin{equation}
{\bf{\hat A}}({{\bf{R}}_o}) \cdot \,{{{\bf{\hat P}}}_R,t} =  - {\bf{\hat A}}({{\bf{R}}_o},t) \cdot \left[ {\frac{1}{R}{\bf{n}} \times {\bf{N}} + i{\bf{n}}\frac{\partial }{{\partial R}}} \right]
\label{FieldCoupling3}
\end{equation}
which shows that the radiation field couples to both molecular rotations and molecular vibrations.  The first term in Eq.~(\ref{FieldCoupling3}) is a coupling of the radiation field to molecular rotations, through the molecular angular momentum ${\bf N}$. The second term in Eq.~(\ref{FieldCoupling3}) is a coupling of the radiation field to molecular vibrations through the length of the inter-nuclear distance $R$.

Babiker et al.\ investigated the interaction of a hydrogenic bound system consisting of a nucleus and a bound electron, interacting with a linearly polarized radiation field carrying  orbital angular momentum~\cite{Babiker2002}. They found that an exchange of orbital angular momentum in an electric dipole transition occurs only between light and the center of mass motion of the atom---they concluded that the internal ''electronic-type'' motion does not participate in any exchange of orbital angular momentum in a dipole transition.   The presence of the second term in Eq.~({\ref{InteractionHamCMofNucleiiCoords}) suggests that this is not the case for a diatomic molecule.

Alexandrescu et al.  studied a charged diatomic molecule (an ion, with two nuclei) interacting with a Laguerre-Gaussian beam carrying orbital angular momentum~\cite{Alexandrescu2006}.  They seem to disagree with the conclusions of  Babiker et al., stating that the electronic motion and the electromagnetic field can exchange one unit of OAM within the electronic dipole interaction.

I note that the interaction Hamiltonian in Eq.~(\ref{InteractionHamCMofNucleiiCoords}) is not of the form $-{\bf d} \cdot {\bf E}$, which is a dipole coupling to the radiation field ${\bf E}$.  The interaction Hamiltonian in  Eq.~(\ref{InteractionHamCMofNucleiiCoords}) contains all multipole couplings with the radiation field.    The  interaction Hamiltonian in Eq.~(\ref{InteractionHamCMofNucleiiCoords}) allows investigation of transfer of angular momentum from the radiation field to the molecule, including transfer of angular momentum to the rotation of the nuclei.  Regarding the dipole moment, consider the dipole moment of the molecule, expressed in terms of variables ${\bf R''}_i$ and ${\bf R''}_\alpha $ in the center-of-mass-of-nuclei coordinates:
\begin{equation}
{\bf{d}} =  - {\bf{e}}\sum\limits_{i = 1}^n {{{{\bf{R''}}}_i}}  + e\sum\limits_{\alpha  = 1}^2 {{Z_\alpha }{{{\bf{R''}}}_\alpha }} 
\label{dipoleMoment}
\end{equation}
When the dipole moment is transformed to the set of coordinates $\{ {\bf R}_o, {\bf R}, {\bf R''}_1, \ldots  {\bf R''}_n \}$, it becomes
\begin{eqnarray}
{\bf{d}} &  = &  - {\bf{e}}\left[ {1 + \frac{m}{M}\left( {{Z_1} + {Z_2}} \right)} \right]\sum\limits_{i = 1}^n {{{{\bf{R''}}}_i}}  \nonumber \\ 
          & +  & e\left( {{Z_1} + {Z_2}} \right){{\bf{R}}_o} - e\left( {\frac{{{Z_1}}}{{{M_1}}} - \frac{{{Z_2}}}{{{M_2}}}} \right)\mu {\kern 1pt} {\bf{R}}
\label{DipoleMomentCMnucleiCoordinates}
\end{eqnarray}
In order to isolate the rotational part of the Hamiltonian, contained in ${\bf P}_{\bf R}$, it was necessary to transform to coordinates $\{ {\bf R}_o, {\bf R}, {\bf R''}_1, \ldots  {\bf R''}_n \}$.  However, after performing his transformation, the Hamiltonian in Eq.~(\ref{InteractionHamCMofNucleiiCoords}) is not of the form of a dipole coupling to the radiation field, $-{\bf d} \cdot {\bf E}$.  This is not a result of the transformation itself, rather it is due to the added rotational degrees of freedom of a molecule as compared to an atom.

\section{\label{TransitionProbabilities}Transition Probabilities}
The form of the molecule-radiation interaction in Eq.~(\ref{InteractionHamCMofNucleiiCoords}) shows that there is a clear coupling of the radiation field to molecular rotations. The center of mass coordinate, ${\bf R}_o$, plays a central role in this coupling. As stated earlier, in order to determine the effect of this coupling on observable quantities, transition matrix elements and probabilities for photon absorption must be calculated.  In order to estimate these matrix elements, we must have knowledge of the form of the wavefunctions for the diatomic molecule, which are schematically shown in Eq.~(\ref{wavefunction}). We must also assume some form for the modes of the radiation field, for example, vector spherical harmonics or Laguerre-Gaussian modes.  Group theoretic methods based on tensor operators can then be used to determine ratios of transition matrix elements~\cite{Judd1975,Zare1988,Brown2003}.  This work is in progress. However, some statements can be made about parity selection rules and how the transition matrix elements scale.

Exact selection rules based on conservation of parity can be written down. Assume that under inversion of coordinates (under the parity operator), the parity of the initial and final molecular wavefunctions are $P_i$ and $P_f$, respectively, and the parity of the emitted photon is $P_\gamma$, where parity eigenvalues takes values $\pm 1$. If the initial state of the whole system has one photon, and the final state has no photons, then parity conservation requires $P_i \, P_\gamma =  P_f$, or, since each parity eigenvalue is $\pm 1$, parity conservation requires~\cite{LL_QuantumElectrodynamics}
\begin{equation}
P_i \, P_f = P_\gamma 
\label{ParityConservation}
\end{equation}
Electric $Ej$ photons have parity $P^{Ej}_\gamma=(-1)^j$ and magnetic $Mj$  photons have parity  $P^{Mj}_\gamma=(-1)^{j+1}$.  For absorbing one photon of a given type, parity conservation requires 
\begin{equation}
P_i \, P_f = (-1)^j   \qquad   {\rm electric \,\,\, {\it Ej} \,\,\,  photon}
\label{ParityConservationElectric}
\end{equation}
or 
\begin{equation}
P_i \, P_f = (-1)^{j+1} \qquad   {\rm magnetic \,\,\, {\it Mj} \,\,\, photon}
\label{ParityConservationMagnetic}
\end{equation}

An estimate of the scaling of transition probabilities can be made by considering the molecule as a system of charges  interacting with a radiation field.  Consider a molecule in an initial state with energy $E_i$,  absorbing a single photon of energy $\hbar \omega_k = \hbar c k$, and making a transition to a final state with energy $E_f$.  The transition matrix element contains the quantity $k r$, which  enters in the vector potential in Eqs.~(\ref{PhotonMagneticWaveFunction}) and~(\ref{PhotonElectricWaveFunction}). The quantity $r$ has a characteristic scale of a diatomic molecule, on the order of $a\sim 10^{-10}$ m.   For  transitions between rotational states,  I take the frequency $f \sim 300$ GHz.  Therefore, there is a small dimensionless quantity, $k r \sim k a \sim 2 \pi  a / \lambda  \sim 6 \times 10^{-7}$ in the transition matrix elements.    The vector potentials in Eqs.~(\ref{PhotonMagneticWaveFunction}) and~(\ref{PhotonElectricWaveFunction}) depend on the function $f_l( k r)$, which in turn depend on the Bessel function $J_{l +\frac{1}{2}} ( k r )$. Using the small argument, $ kr \ll 1$, expansion of the Bessel function,
in Eq.~(\ref{RadialFunctions}), the radial function, $f_l( k r)$, scales as (up to a constant):
\begin{equation}
{f_l}\left( {k\,r} \right) \sim {\left( {k{\kern 1pt} r} \right)^{l + \frac{1}{2}}}\frac{1}{{1 \cdot 3 \cdot 5 \cdots (2l + 1)}}
\label{RadialFnScaling}
\end{equation}
The vector potential for $Ej$ photons scales as \mbox{${\bf{A}}_{kjm}^E\left( {\bf{r}} \right) \sim {f_{j - 1}}\left( {k\,r} \right)$}, and the vector potential for $Mj$ photons scales as \mbox{${\bf{A}}_{kjm}^M \left( {\bf{r}} \right) \sim {f_j}\left( {k\,r} \right)$}. The probability of absorbing a photon scales as the square of the vector potential (square of the matrix element).  Therefore,  the ratio of the probability for absorbing an $Mj$ photon, $P(Mj)$, divided by the probability of absorbing an $Ej$ photon, $P(Ej)$, scales as 
\begin{equation}
\frac{{P\left( {Mj} \right)}}{{P\left( {Ej} \right)}} \sim \frac{{{{\left( {k{\kern 1pt} a} \right)}^2}}}{{{{\left( {2j + 1} \right)}^2}}} \sim \frac{{{{4 \times 10}^{ - 13}}}}{{{{\left( {2j + 1} \right)}^2}}}
\label{ProbMj/EjRatioScaling}
\end{equation}
For a given angular momentum $j$, the probability of absorption of a magnetic photon is much less than absorption of an electric photon. 

The ratio of the probability for absorbing an $E(j+1)$ photon, $P(E(j+1)$, divided by the probability of absorbing an $Ej$ photon, $P(Ej)$, also scales as 
\begin{equation}
\frac{{P\left( {E(j + 1)} \right)}}{{P\left( {Ej} \right)}} \sim \frac{{{{\left( {k{\kern 1pt} a} \right)}^2}}}{{{{\left( {2j + 1} \right)}^2}}} \sim \frac{{{{4\times 10}^{ - 13}}}}{{{{\left( {2j + 1} \right)}^2}}}
\label{ProbE(j+1)/EjRatioScaling}
\end{equation}
Form Eq.~(\ref{ProbE(j+1)/EjRatioScaling}), we see that the probability  of absorbing higher angular momentum photons rapidly decreases with increasing $j$. As the simplest example, the ratio of probability of absorbing an $E2$ photon to the probability of absorbing an $E1$ photon is: 
\begin{equation}
\frac{{P\left( {E2} \right)}}{{P\left( {E1} \right)}} \sim {4 \times 10^{ - 14}}
\label{ProbE2/Prob1E_RatioScaling}
\end{equation}

The scaling of the absorption probabilities for molecules, given in Eqs.~(\ref{ProbMj/EjRatioScaling})--(\ref{ProbE2/Prob1E_RatioScaling}), makes it a challenging proposition to observe absorption of photons having angular momentum $j>1$.

\section{\label{Summary}Summary}

I have investigated the interaction between a diatomic molecule and photons with well-defined angular momentum. I have exploited the  transformation from laboratory coordinates to coordinates with origin at the center of mass of the nuclei to show the nuclear rotational degrees of freedom explicitly (through the vector ${\bf R}$).    For the photon modes, I have assumed spherical vector harmonics in a large sphere of radius $R$. These modes are exact solutions of Maxwell equations, and consequently, any other modes, such as the Laguerre-Gaussian modes, can be expanded in terms of them. However, in general, spin and orbital angular momentum are not separable in vector spherical harmonic modes.  I used the standard minimal coupling between elementary charges and radiation field to couple the radiation field to the molecule, neglecting quadratic terms in the vector potential and neglecting relativistic interaction with spins.  For molecules, which absorb at millimeter wavelengths, there is a small parameter, $ k \, a$ where $k$ is a wave vector and $a$ is the size of the molecule.  This small parameter was used to determine the scaling of absorption probabilities of $Ej$ and $Mj$ type photons.  For transitions between rotational levels of the molecule having $\Delta J>1$, the probability  of absorption of photons with angular momentum $j>1$ is small, because the factor $k a  \ll 1$, where wave vector $k = 2 \pi/\lambda $ corresponds to photons in the millimeter wavelength region,  see Section~\ref{TransitionProbabilities}.  Note that the small parameter, $ k \, a$, also exists for atoms~\cite{Huang1994,Jauregui2004,Grinter2008}, however, it is larger for atoms because $k$ corresponds to visible wavelengths, whereas for rotational transitions in molecules $k$ corresponds to millimeter wavelengths. 

For electric photons having arbitrary angular momentum $j \hbar$, the probability of rotational transition by absorbing an electric $E(j+1)$ photon divided by the probability of absorbing an electric $Ej$ photon, scales as $ (k a)^2 /( 2 j+1)^2$. The probability of absorbing a magnetic $Mj$ photon is much smaller than the probability of absorbing an electric  $Ej$ photon.  The probability of absorbing a magnetic $Mj$ photon, divided by the probability of absorbing an electric $Ej$ photon, also scales as $ (k a)^2 /( 2 j+1)^2$.

\begin{acknowledgments}
The author acknowledges interesting discussions with Henry Everitt and Matt Goodman.  
\end{acknowledgments}

%
\bibliographystyle{apsrev}
\bibliography{References-Quantum}

\begin{thebibliography}{30}
\expandafter\ifx\csname natexlab\endcsname\relax\def\natexlab#1{#1}\fi
\expandafter\ifx\csname bibnamefont\endcsname\relax
  \def\bibnamefont#1{#1}\fi
\expandafter\ifx\csname bibfnamefont\endcsname\relax
  \def\bibfnamefont#1{#1}\fi
\expandafter\ifx\csname citenamefont\endcsname\relax
  \def\citenamefont#1{#1}\fi
\expandafter\ifx\csname url\endcsname\relax
  \def\url#1{\texttt{#1}}\fi
\expandafter\ifx\csname urlprefix\endcsname\relax\def\urlprefix{URL }\fi
\providecommand{\bibinfo}[2]{#2}
\providecommand{\eprint}[2][]{\url{#2}}

\bibitem[{\citenamefont{Born and Oppenheimer}(1927)}]{Born1927}
\bibinfo{author}{\bibfnamefont{M.}~\bibnamefont{Born}} \bibnamefont{and}
  \bibinfo{author}{\bibfnamefont{J.~R.} \bibnamefont{Oppenheimer}},
  \bibinfo{journal}{Ann. Phys.} \textbf{\bibinfo{volume}{84}},
  \bibinfo{pages}{457–484} (\bibinfo{year}{1927}).

\bibitem[{\citenamefont{Wang}(1929)}]{Wang1929}
\bibinfo{author}{\bibfnamefont{S.~C.} \bibnamefont{Wang}},
  \bibinfo{journal}{Phys. Rev.} \textbf{\bibinfo{volume}{34}},
  \bibinfo{pages}{243} (\bibinfo{year}{1929}).

\bibitem[{\citenamefont{Landau and Lifshitz}(1984)}]{LL_QuantumMechanics}
\bibinfo{author}{\bibfnamefont{L.~D.} \bibnamefont{Landau}} \bibnamefont{and}
  \bibinfo{author}{\bibfnamefont{E.~M.} \bibnamefont{Lifshitz}},
  \emph{\bibinfo{title}{QuantumMechanics--Non-Relativistic Theory}}
  (\bibinfo{publisher}{Pergamon Press}, \bibinfo{address}{New York, N.Y. USA},
  \bibinfo{year}{1984}), \bibinfo{edition}{3rd} ed.

\bibitem[{\citenamefont{Davydov}(1965)}]{Davydov_QuantumMechanics}
\bibinfo{author}{\bibfnamefont{A.~S.} \bibnamefont{Davydov}},
  \emph{\bibinfo{title}{Quantum Mechanics}} (\bibinfo{publisher}{Pergamon
  Press}, \bibinfo{address}{New York, N.Y., USA}, \bibinfo{year}{1965}),
  \bibinfo{edition}{2nd} ed.

\bibitem[{\citenamefont{Van~Vleck}(1929)}]{VanVleck1929}
\bibinfo{author}{\bibfnamefont{J.~H.} \bibnamefont{Van~Vleck}},
  \bibinfo{journal}{Phys. Rev.} \textbf{\bibinfo{volume}{33}},
  \bibinfo{pages}{467} (\bibinfo{year}{1929}),
  \urlprefix\url{http://link.aps.org/doi/10.1103/PhysRev.33.467}.

\bibitem[{\citenamefont{Meyer}(2002)}]{Meyer2002}
\bibinfo{author}{\bibfnamefont{H.}~\bibnamefont{Meyer}},
  \bibinfo{journal}{Annu. Rev. Phys. Chem.} \textbf{\bibinfo{volume}{53}},
  \bibinfo{pages}{141} (\bibinfo{year}{2002}).

\bibitem[{\citenamefont{Brown and Carrington}(2003)}]{Brown2003}
\bibinfo{author}{\bibfnamefont{J.}~\bibnamefont{Brown}} \bibnamefont{and}
  \bibinfo{author}{\bibfnamefont{A.}~\bibnamefont{Carrington}},
  \emph{\bibinfo{title}{Rotational Spectroscopy of Diatomic Molecules}}
  (\bibinfo{publisher}{Cambridge University Press}, \bibinfo{address}{New York,
  N. Y., USA}, \bibinfo{year}{2003}).

\bibitem[{\citenamefont{Akheizer and
  Berestetskii}(1965)}]{AB_QuantumElectrodynamics}
\bibinfo{author}{\bibfnamefont{A.~I.} \bibnamefont{Akheizer}} \bibnamefont{and}
  \bibinfo{author}{\bibfnamefont{V.~B.} \bibnamefont{Berestetskii}},
  \emph{\bibinfo{title}{Quantum Electrodynamics}} (\bibinfo{publisher}{John
  Wiley \& Sons}, \bibinfo{address}{New York, N.Y., USA},
  \bibinfo{year}{1965}).

\bibitem[{\citenamefont{Berestetskii et~al.}(1982)\citenamefont{Berestetskii,
  Lifshitz, and Pitaevskii}}]{LL_QuantumElectrodynamics}
\bibinfo{author}{\bibfnamefont{V.~B.} \bibnamefont{Berestetskii}},
  \bibinfo{author}{\bibfnamefont{E.~M.} \bibnamefont{Lifshitz}},
  \bibnamefont{and} \bibinfo{author}{\bibfnamefont{L.~P.}
  \bibnamefont{Pitaevskii}}, \emph{\bibinfo{title}{Quantum Electrodynamics}}
  (\bibinfo{publisher}{Pergamon Press}, \bibinfo{address}{Elmsford, New York,
  USA}, \bibinfo{year}{1982}), \bibinfo{edition}{2nd} ed.

\bibitem[{\citenamefont{Atkins and Woolley}(1970)}]{Atkins1970}
\bibinfo{author}{\bibfnamefont{P.~W.} \bibnamefont{Atkins}} \bibnamefont{and}
  \bibinfo{author}{\bibfnamefont{R.~G.} \bibnamefont{Woolley}},
  \bibinfo{journal}{Proc. Roy. Soc. Lond. A} \textbf{\bibinfo{volume}{319}},
  \bibinfo{pages}{549} (\bibinfo{year}{1970}).

\bibitem[{\citenamefont{Woolley}(1971)}]{Woolley1971}
\bibinfo{author}{\bibfnamefont{R.~G.} \bibnamefont{Woolley}},
  \bibinfo{journal}{Proc. Roy. Soc. Lond. A} \textbf{\bibinfo{volume}{321}},
  \bibinfo{pages}{557} (\bibinfo{year}{1971}).

\bibitem[{\citenamefont{Woolley}(1975)}]{Woolley1975}
\bibinfo{author}{\bibfnamefont{R.~G.} \bibnamefont{Woolley}}, in
  \emph{\bibinfo{booktitle}{Volume 33}}, edited by
  \bibinfo{editor}{\bibfnamefont{I.}~\bibnamefont{Progogine}} \bibnamefont{and}
  \bibinfo{editor}{\bibfnamefont{S.~A.} \bibnamefont{Rice}}
  (\bibinfo{publisher}{University of Chicago}, \bibinfo{address}{Chicago,
  Illinois}, \bibinfo{year}{1975}), Advances in Chemical Physics, pp.
  \bibinfo{pages}{153--233}.

\bibitem[{\citenamefont{Craig and
  Thirunamachandran}(1998)}]{Craig_MolecularQuantumElectrodynamics}
\bibinfo{author}{\bibfnamefont{D.~P.} \bibnamefont{Craig}} \bibnamefont{and}
  \bibinfo{author}{\bibfnamefont{T.}~\bibnamefont{Thirunamachandran}},
  \emph{\bibinfo{title}{Molecular Quantum Electrodynamics}}
  (\bibinfo{publisher}{Dover Publications, Inc.}, \bibinfo{address}{Mineola,
  N.Y., USA}, \bibinfo{year}{1998}), \bibinfo{edition}{dover edition} ed.

\bibitem[{\citenamefont{Sindelka and Moiseyev}(2006)}]{Sindelka2006}
\bibinfo{author}{\bibfnamefont{M.}~\bibnamefont{Sindelka}} \bibnamefont{and}
  \bibinfo{author}{\bibfnamefont{N.}~\bibnamefont{Moiseyev}},
  \bibinfo{journal}{J. Phys. Chem. A} \textbf{\bibinfo{volume}{110}},
  \bibinfo{pages}{5561–5571} (\bibinfo{year}{2006}),
  \urlprefix\url{http://arxiv.org/abs/quant-ph/0601195}.

\bibitem[{\citenamefont{Beth}(1936)}]{Beth1936}
\bibinfo{author}{\bibfnamefont{R.~A.} \bibnamefont{Beth}},
  \bibinfo{journal}{Phys. Rev.} \textbf{\bibinfo{volume}{50}},
  \bibinfo{pages}{115} (\bibinfo{year}{1936}).

\bibitem[{\citenamefont{Allen et~al.}(1992)\citenamefont{Allen, Beijersbergen,
  Spreeuw, and Woerdman}}]{Allen1992}
\bibinfo{author}{\bibfnamefont{L.}~\bibnamefont{Allen}},
  \bibinfo{author}{\bibfnamefont{M.~W.} \bibnamefont{Beijersbergen}},
  \bibinfo{author}{\bibfnamefont{R.~J.~C.} \bibnamefont{Spreeuw}},
  \bibnamefont{and} \bibinfo{author}{\bibfnamefont{J.~P.}
  \bibnamefont{Woerdman}}, \bibinfo{journal}{Phys. Rev. A}
  \textbf{\bibinfo{volume}{45}}, \bibinfo{pages}{8185} (\bibinfo{year}{1992}),
  \urlprefix\url{http://link.aps.org/doi/10.1103/PhysRevA.45.8185}.

\bibitem[{\citenamefont{Torres and Torner}(2011)}]{TwistedPhotonsBook}
\bibinfo{editor}{\bibfnamefont{J.~P.} \bibnamefont{Torres}} \bibnamefont{and}
  \bibinfo{editor}{\bibfnamefont{L.}~\bibnamefont{Torner}}, eds.,
  \emph{\bibinfo{title}{Twisted Photons: Applications of Light With Orbital
  Angular Momentum}} (\bibinfo{publisher}{WILEY-VCH Verlag GmbH \& Co. KGaA},
  \bibinfo{address}{Weinheim, Germany}, \bibinfo{year}{2011}).

\bibitem[{\citenamefont{Allen et~al.}(1999)\citenamefont{Allen, Padgett, and
  Babiker}}]{Allen1999}
\bibinfo{author}{\bibfnamefont{L.}~\bibnamefont{Allen}},
  \bibinfo{author}{\bibfnamefont{M.~J.} \bibnamefont{Padgett}},
  \bibnamefont{and} \bibinfo{author}{\bibfnamefont{M.}~\bibnamefont{Babiker}},
  in \emph{\bibinfo{booktitle}{Progress on Optics, Volume 39}}, edited by
  \bibinfo{editor}{\bibfnamefont{E.}~\bibnamefont{Wolf}}
  (\bibinfo{publisher}{Elsevier}, \bibinfo{address}{New York, N.Y. USA},
  \bibinfo{year}{1999}), Progress on Optics.

\bibitem[{\citenamefont{Huang}(1994)}]{Huang1994}
\bibinfo{author}{\bibfnamefont{K.-N.} \bibnamefont{Huang}},
  \bibinfo{journal}{Am. J. Phys.} \textbf{\bibinfo{volume}{62}},
  \bibinfo{pages}{73} (\bibinfo{year}{1994}).

\bibitem[{\citenamefont{J\'auregui}(2004)}]{Jauregui2004}
\bibinfo{author}{\bibfnamefont{R.}~\bibnamefont{J\'auregui}},
  \bibinfo{journal}{Phys. Rev. A} \textbf{\bibinfo{volume}{70}},
  \bibinfo{pages}{033415} (\bibinfo{year}{2004}),
  \urlprefix\url{http://link.aps.org/doi/10.1103/PhysRevA.70.033415}.

\bibitem[{\citenamefont{Grinter}(2008)}]{Grinter2008}
\bibinfo{author}{\bibfnamefont{R.}~\bibnamefont{Grinter}}, \bibinfo{journal}{J.
  Phys. B: At. Mol. Opt. Phys.} \textbf{\bibinfo{volume}{41}},
  \bibinfo{pages}{1} (\bibinfo{year}{2008}).

\bibitem[{\citenamefont{Thomas}(1970{\natexlab{a}})}]{Thomas1970}
\bibinfo{author}{\bibfnamefont{I.~L.} \bibnamefont{Thomas}},
  \bibinfo{journal}{Phys. Rev. A} \textbf{\bibinfo{volume}{2}},
  \bibinfo{pages}{72} (\bibinfo{year}{1970}{\natexlab{a}}),
  \urlprefix\url{http://link.aps.org/doi/10.1103/PhysRevA.2.72}.

\bibitem[{\citenamefont{Thomas and Joy}(1970)}]{Thomas1970a}
\bibinfo{author}{\bibfnamefont{I.~L.} \bibnamefont{Thomas}} \bibnamefont{and}
  \bibinfo{author}{\bibfnamefont{H.~W.} \bibnamefont{Joy}},
  \bibinfo{journal}{Phys. Rev. A} \textbf{\bibinfo{volume}{2}},
  \bibinfo{pages}{1200} (\bibinfo{year}{1970}),
  \urlprefix\url{http://link.aps.org/doi/10.1103/PhysRevA.2.1200}.

\bibitem[{\citenamefont{Thomas}(1970{\natexlab{b}})}]{Thomas1970b}
\bibinfo{author}{\bibfnamefont{I.~L.} \bibnamefont{Thomas}},
  \bibinfo{journal}{Phys. Rev. A} \textbf{\bibinfo{volume}{2}},
  \bibinfo{pages}{1675} (\bibinfo{year}{1970}{\natexlab{b}}),
  \urlprefix\url{http://link.aps.org/doi/10.1103/PhysRevA.2.1675}.

\bibitem[{\citenamefont{Thomas}(1971)}]{Thomas1971}
\bibinfo{author}{\bibfnamefont{I.~L.} \bibnamefont{Thomas}},
  \bibinfo{journal}{Phys. Rev. A} \textbf{\bibinfo{volume}{3}},
  \bibinfo{pages}{1022} (\bibinfo{year}{1971}),
  \urlprefix\url{http://link.aps.org/doi/10.1103/PhysRevA.3.1022}.

\bibitem[{\citenamefont{Romero et~al.}(2002)\citenamefont{Romero, Andrews, and
  Babiker}}]{Romero2002}
\bibinfo{author}{\bibfnamefont{L.~C.~D.} \bibnamefont{Romero}},
  \bibinfo{author}{\bibfnamefont{D.~L.} \bibnamefont{Andrews}},
  \bibnamefont{and} \bibinfo{author}{\bibfnamefont{M.}~\bibnamefont{Babiker}},
  \bibinfo{journal}{J. Opt. B: Quantum Semiclass. Opt.}
  \textbf{\bibinfo{volume}{4}}, \bibinfo{pages}{S66} (\bibinfo{year}{2002}).

\bibitem[{\citenamefont{Judd}(1975)}]{Judd1975}
\bibinfo{author}{\bibfnamefont{B.~R.} \bibnamefont{Judd}},
  \emph{\bibinfo{title}{Angular Momentum Theory for Diatomic Molecules}}
  (\bibinfo{publisher}{Academic Press}, \bibinfo{address}{New York, USA},
  \bibinfo{year}{1975}).

\bibitem[{\citenamefont{Zare}(1988)}]{Zare1988}
\bibinfo{author}{\bibfnamefont{R.~N.} \bibnamefont{Zare}},
  \emph{\bibinfo{title}{Angular Momentum: Understanding Spatial Aspecta in
  Chemistry and Physics}} (\bibinfo{publisher}{Wiley-Interscience Publication},
  \bibinfo{address}{New York, USA}, \bibinfo{year}{1988}).

\bibitem[{\citenamefont{Babiker et~al.}(2002)\citenamefont{Babiker, Bennett,
  Andrews, and D\'avila~Romero}}]{Babiker2002}
\bibinfo{author}{\bibfnamefont{M.}~\bibnamefont{Babiker}},
  \bibinfo{author}{\bibfnamefont{C.~R.} \bibnamefont{Bennett}},
  \bibinfo{author}{\bibfnamefont{D.~L.} \bibnamefont{Andrews}},
  \bibnamefont{and} \bibinfo{author}{\bibfnamefont{L.~C.}
  \bibnamefont{D\'avila~Romero}}, \bibinfo{journal}{Phys. Rev. Lett.}
  \textbf{\bibinfo{volume}{89}}, \bibinfo{pages}{143601}
  (\bibinfo{year}{2002}),
  \urlprefix\url{http://link.aps.org/doi/10.1103/PhysRevLett.89.143601}.

\bibitem[{\citenamefont{Alexandrescu et~al.}(2006)\citenamefont{Alexandrescu,
  Cojoc, and Fabrizio}}]{Alexandrescu2006}
\bibinfo{author}{\bibfnamefont{A.}~\bibnamefont{Alexandrescu}},
  \bibinfo{author}{\bibfnamefont{D.}~\bibnamefont{Cojoc}}, \bibnamefont{and}
  \bibinfo{author}{\bibfnamefont{E.~D.} \bibnamefont{Fabrizio}},
  \bibinfo{journal}{Phys. Rev. Lett.} \textbf{\bibinfo{volume}{96}},
  \bibinfo{pages}{243001} (\bibinfo{year}{2006}).

\end{thebibliography}
%
\end{document}